\documentclass{aastex}
\usepackage{spr-astr-addons}
\usepackage{url}
\RequirePackage{color}
\usepackage{xcolor}

\newcommand{\at}{Alessi-Teutsch~9}

\begin{document}

\title{Study of the open cluster Alessi-Teutsch 9 (ASCC~10)\\
       using multiband photometry and \textbf{\textit{Gaia}} EDR3}

\shorttitle{Open cluster Alessi-Teutsch~9}
\shortauthors{Sanchez et al.}

\author{N\'estor S\'anchez\altaffilmark{1}} 
\affil{nestor.sanchezd@campusviu.es} \and
\author{F\'atima L\'opez-Mart\'inez\altaffilmark{2}} \and
\author{Sandra Ocando\altaffilmark{3}} \and
\author{Pere Blay\altaffilmark{1}}

\altaffiltext{1}{Universidad Internacional de Valencia (VIU),
                 C/Pintor Sorolla 21, 46002, Valencia, Spain.}
\altaffiltext{2}{Centro de Estudios de F\'isica del Cosmos de
                 Arag\'on (CEFCA), Unidad Asociada al CSIC, 
                 Plaza de San Juan 1, 44001, Teruel, Spain.}
\altaffiltext{3}{Instituto de Astrof\'isica de Andaluc\'ia, 
                 CSIC, Glorieta de la Astronom\'ia S/N, Granada,
                 18008, Spain.}

\begin{abstract}
There is a growing interest in the automated characterization
of open clusters using data from the {\it Gaia} mission. This
work evidences the importance of choosing an appropriate
sampling
radius (the radius of the circular region
around the cluster used to extract the data)
and the usefulness of additional multiband
photometry in order to achieve accurate results. We address
this issue using as a case study the cluster Alessi-Teutsch~9.
The optimal sampling is determined by counting the number of
assigned members at different sampling radii. By using this
strategy with data from {\it Gaia} EDR3 and with observed
photometry in $12$ bands
spanning the optical range from $3000$ to
$10000$ \AA, approximately,
we are able to obtain reliable
members and to determine the properties of the cluster.
The spatial distribution of stars show a two-component
structure with a central core of radius $\sim 12-13$
arcmin and an outer halo extending out to $35$ arcmin.
With the derived cluster distance ($654$ pc) we obtain that
the number density of stars is $\simeq 0.06$ star/pc$^3$,
making Alessi-Teutsch~9 one of the less dense known open
clusters. The short relaxation time reveals that it is a
dynamically relaxed and gravitationally bound system.
\end{abstract}

\keywords{methods: data analysis --
          open clusters and associations: general --
          open clusters and associations: individual: Alessi-Teutsch~9}

\section{Introduction}

Open clusters (OCs) are very important components of our 
Galaxy. A detailed knowledge of their properties, such as 
distance, age, metallicity or reddening is necessary for a 
proper understanding of the structure and evolution of the 
Milky Way \citep[see e.g. the review by][]{Kru19}. With the 
advent of the ESA’s {\it Gaia} space mission \citep{Gai16}, its 
second data release \citep[DR2,][]{Gai18} and, more recently, 
the Early Data Release 3 \citep[EDR3,][]{Gai20}, astronomers 
have an homogeneous source of data with unprecedented 
astrometric precision and accuracy that have allowed to 
increase the census 
of
known OCs and to improve the 
determinations of their properties. In the last years, 
the number of papers using data from {\it Gaia} and different 
techniques (from simple visual inspection to supervised 
or unsupervised machine learning techniques) and reporting 
the discovery of new OCs has increased notoriously 
\citep{Cas18,Can19,Fer19,Liu19,Sim19,Cas20,Casado21,He21,Fer21,Hun21,Xia21}.
An important requirement to be able to carry out reliable 
studies of the Galactic cluster population is that OC 
parameters are derived in a homogeneous manner. The 
inhomogeneous analysis combining different data sets 
and/or methods may lead to discrepant or biased results, 
as has been noted by \citet{Net15} and \citet{Car17}.
In this sense, the high quality of the {\it Gaia} data
allows the systematic and homogeneous derivation of OC
parameters with very good precision
\citep[e.g.][]{Sou18,Bos19,Can20a,Can20b,Aga21,Tar21}.

There are, however, two important issues that must be 
taken into account in deriving cluster parameters, 
especially when performing massive data processing. The 
first one refers to the size of the sample used 
to extract the data to work with. This is a very 
important first step that usually is decided in 
advance to data processing. As discussed in 
\citet{San10}, if the radius of the circular 
area around the cluster used to extract the data 
(the sampling radius, $R_s$) is much smaller than 
the actual cluster radius ($R_c$) then the cluster 
is obviously subsampled. In order to avoid this 
situation, usually a ``large enough" $R_s$ is selected. 
The problem is that if $R_s \gg R_c$ then the contamination 
by field stars affects 
the determination of 
memberships and, consequently, the determination of the cluster 
properties, including the radius itself \citep{San10}.
Ideally a relatively high $R_s$ value
should not affect the membership determination, but
depending on membership assignment criteria and/or
measurement errors the contamination by field stars
could become significant and, in any case, it tends
to be higher as $R_s$ increases.
Without a previous knowledge of an approximate value of 
the cluster radius it is difficult to choose the optimal 
sampling radius $R_s \simeq R_c$. This is probably one of the 
reasons for the large discrepancies in $R_c$ reported in 
the literature because, independently on
the used membership assignment method, 
selected members tend to be spread over the sampled area and 
the a posteriori determination of $R_c$ based, for instance, 
in a radial star density profile, can be biased by the 
previous selection of $R_s$
\citep[see discussions in][]{San10,San18,San20}.
For the case of the OC
presented in this work (\at), previously calculated
radius values range from $25.7$~arcmin \citep{Sam17}
to $31.8$~arcmin \citep{Kha13}, but more discrepant
values can be found for other clusters
\citep[see for instance Fig.~5 in][]{San20}. 

The second relevant consideration relates to the limitation of
using only {\it Gaia} photometry, which may be motivated in part 
by the high quality (and also homogeneity) of the {\it Gaia}
data. However, using only the relatively large width 
{\it Gaia} bands can result in degeneracy problems and/or lead to 
strong systematics when inferring OC parameters \citep{And18}. 
The obvious solution would be to complement {\it Gaia} photometry
with multiband photometry from other surveys such as SDSS
\citep{Yor00} or 2MASS \citep{Skr06} as done, for example,
in \citet{Tad18}. The larger the set of used photometric 
bands, the more accurate the estimated stellar parameters 
without the need of resorting to spectroscopy but, as mentioned 
above, some caution is required when combining several passbands 
from different photometric systems.

Based on these considerations, we have started a project to 
observe and characterize OCs using the 80~cm JAST80 
telescope at the Observatorio Astrof\'isico de Javalambre 
(OAJ) in Teruel, Spain. The telescope has a panoramic camera 
(T80Cam) with a large-format CCD of $9200 \times 9200$ pixels 
providing an effective field of view of $1.4 \times 1.4$ deg$^2$ 
that is particularly useful for wide field OCs. Moreover, it 
hosts a set of $12$ optical filters, originally defined for 
the J-PLUS survey \citep{Cen19}, spanning the entire optical 
range. Here we report the results of the pilot campaign 
for the cluster \at\ \citep{Dia02,Kha05}, also known as ASCC~10 
and MWSC~275 \citep{Bic19}. \at, located in the second Galactic 
quadrant ($l = 155.61$ deg, $b=-17.79$ deg) \citep{Kha05}, is a 
relatively large cluster for which apparent radii in the range 
$\sim 25-32$ arcmin have been reported in works previous to the 
{\it Gaia} era \citep{Kha05,Kha13,Dia14,Sam17}. The mean cluster 
proper motion, based on less precise catalogues as PPMXL 
\citep{Roe10} or UCAC4 \citep{Zac13}, had disagreeing values 
in the ranges $\sim (-3.4)\ -\ (+1.7)$ mas~yr$^{-1}$ in right 
ascension and $\sim (-2.2)\ -\ (-1.0)$ mas~yr$^{-1}$ in 
declination \citep{Kha05,Kha13,Dia14,Sam17,Can18a} and, 
consequently, the number of assigned members have varied 
in the range $\sim 10^2-10^3$. More recently and using data 
from {\it Gaia} DR2, \citet{Can18b} \citep[see also][]{Can20a,Can20b}
located the cluster proper motion centroid at $(-1.737,-1.368)$ 
mas~yr$^{-1}$ with a total of $71$ probable members. The 
method used by \citet{Can20a} applies $k$-means clustering 
to search for groupings in proper motions and parallaxes. 
As other authors, a relatively high initial sampling radius 
was intentionally assumed and, even though they reported 
that the radius containing half the members is $33.5$ arcmin,
the assigned cluster members spread over a radius of $\sim 70$
arcmin.

In this work we combined $12$-band photometry from OAJ 
for \at\ with {\it Gaia} EDR3 data to show (1) the importance 
of a {\it previous} identification of the optimal sampling 
radius and (2) the advantage of incorporate multiband 
photometry for a precise and accurate estimation of OC 
properties. Section~\ref{sec_data} describes the observed
and used data sets. In Section~\ref{sec_sampling} we
determine the optimal sampling radius, which is used in
Section~\ref{sec_members} to determine reliable membership
criteria and to select the final list of cluster members.
Spectral energy distributions for the members, based on
OAJ photometry, are derived in Section~\ref{sec_seds}.
The cluster properties are presented and discussed in
Section~\ref{sec_properties} and, finally, the main
conclusions are summarized in Section~\ref{sec_conclusions}.

\section{Observations and data}
\label{sec_data}

\subsection{OAJ}

Observations were made on February 23 and 24, 2020, with 
the T80Cam panoramic camera attached to the JAST80 
telescope at OAJ \citep{oaj}. The total field of view is 
$1.4 \times 1.4$ deg$^2$ and the pixel scale is $0.55$ 
arcsec/pixel. The set of filters covers the optical range, 
from 3500 to 10000 \AA\ \citep[see][]{Cen19}, and they are
presented in Table~\ref{tabOAJ}, indicating the central
and Full Width Half Maximum ($FWHM$). The log of the
observations (number of exposures $N_{exp}$ and total
exposure time $T_{exp}$) is also provided in Table~\ref{tabOAJ}.
\begin{table}[t]
\caption{List of filters and log of observations for \at.}
\label{tabOAJ}
\begin{tabular}{@{}ccccc}
\hline
Name & $\lambda$ (\AA) & $FWHM$ (\AA) & $N_{epx}$ (s) & $T_{exp}$ (s) \\
\hline
uJAVA & 3485 &  508 & 12 & 603 \\
J0378 & 3785 &  168 & 12 & 813 \\
J0395 & 3950 &  100 & 12 & 843 \\
J0410 & 4100 &  200 & 12 & 408 \\
J0430 & 4300 &  200 & 12 & 378 \\
gSDSS & 4803 & 1409 & 12 &  93 \\
J0515 & 5150 &  200 & 22 & 631 \\
rSDSS & 6254 & 1388 & 18 & 204 \\
J0660 & 6600 &  138 & 12 & 588 \\
iSDSS & 7668 & 1535 & 12 & 138 \\
J0861 & 8610 &  400 & 12 & 978 \\
zSDSS & 9114 & 1409 & 12 & 528 \\
\hline
\end{tabular}
\end{table}
In total there were $160$ images for the cluster \at\ in 
the $12$ available bands. Data reduction and photometric 
calibration were performed by the Data Processing and 
Archiving Unit (UPAD) team \citep{upad,Cen19}.

\subsection{\textbf{\textit{Gaia}} EDR3 data}

We queried the {\it Gaia} EDR3 database around the cluster 
center\footnote{The exact center is irrelevant because 
it is recalculated from the final member selection.} 
at RA(J2000)=51.870 deg and Dec=34.981 deg \citep{Can20a}. 
All the available astrometric and photometric information 
was extracted using different sampling radii according to 
the proposed strategy (see Section~\ref{sec_sampling}). It 
is worth pointing out that we did not apply any magnitude 
cut or filtering on data quality, unlike \citet{Can20a}
that were limited to stars brighter than $G=18$.

\section{Optimal sampling radius}
\label{sec_sampling}

In previous works \citep{San10,San18,San20} we have been 
developing and improving a method for objectively calculating 
the optimal $R_s$ value for OCs. The method is based on the 
behavior of the stars in the proper motion space as $R_s$ 
increases and it is not affected by how its stars are 
spatially distributed. 
For a given $R_s$ value, the first step
is to identify the cluster overdensity in the proper motion
space. For this, radial density profiles are derived for all
the available stars in the proper motion space, that is
assuming each star as the centre of the overdensity. In
each case, an overdensity ``edge" is also determined as the
point at which the profile changes from a steep to a shallow
slope. At this point, average densities are calculated both
for the overdensity (the circular region from the starting
point to the edge) and for the ``local field" (a ring
adjacent to the overdensity region). The ring width is
chosen such that its area is the overdensity area but
with the condition that the minimum number of stars
in the ring is $100$. These calculations are done for
all the starting points (stars) and the cluster overdensity
is identified as the one having the highest density contrast
between the overdense region and the local field.
The core of the method, in its latest version, consists of 
measuring how much the density of stars in the 
cluster
overdensity increases ($\Delta D_{od}$) as $R_s$ 
increases. This change is compared with the change 
in the local field density ($\Delta D_{lf}$), for 
which a ring surrounding the overdensity is used. 
The key point is the assumption that 
$\Delta D_{od}> \Delta D_{lf}$ when $R_s < R_c$ because 
both field stars and cluster stars are being included 
inside the overdensity. However, when $R_s \gtrsim R_c$ 
we expect that $\Delta D_{od} \simeq \Delta D_{lf}$ 
\citep[see details in][]{San20}.

\begin{figure}[t]
\includegraphics[width=\columnwidth]{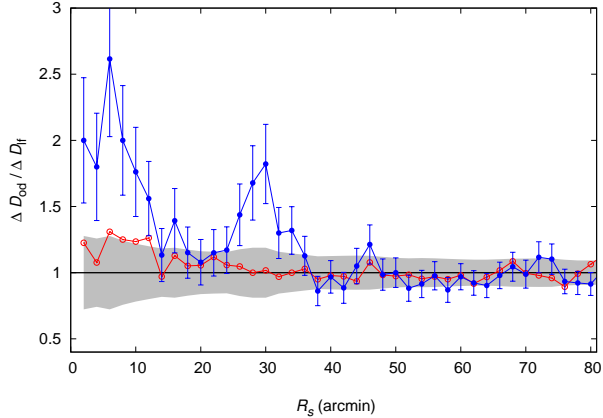}
\caption{Ratio between the density variation in the 
proper motion overdensity ($\Delta D_{od}$) and the 
density variation in the local field region 
($\Delta D_{lf}$) as a function of the sampling 
radius ($R_s$) around the open cluster \at. Red 
symbols refer to the full sample and blue symbols 
to stars having proper motion errors smaller than 
$0.3$ mas~yr$^{-1}$ in right ascension or declination. 
Blue error bars are estimated by assuming Poisson 
statistics whereas the grey area indicates the 
uncertainty associated with local field variations 
(errors for the red symbols are not shown for clarity). 
Black horizontal line, shown as reference, corresponds 
to the case $\Delta D_{od}=\Delta D_{lf}$ expected when 
$R_s$ reaches the actual cluster radius $R_c$.}
\label{fig_eta}
\end{figure}

The application of this method to \at\ using data from 
the {\it Gaia} EDR3 catalog gives the result shown in 
Figure~\ref{fig_eta}.
Direct application to the full data (red symbols) does 
not yield a clear result probably because of the relatively 
low density of stars in \at\ compared to the field stars. 
Normally, this kind of problems can be solved by applying 
some kind of data filtering. At this point it is not 
necessary to use the full sample of stars because the 
goal is not yet to find members but only to make a 
reliable estimate of the cluster radius. If we rule 
out stars with errors in proper motion higher 
than $0.3$ mas~yr$^{-1}$ (the median of the error in 
the {\it Gaia} EDR3 catalog for this sample) then the result 
improves notoriously (blue symbols in Figure~\ref{fig_eta}) 
and the point at which $\Delta D_{od} \simeq \Delta D_{lf}$ 
becomes apparent. The valley in the curve observed around 
$R_s \sim 20$ arcmin could be related to variations in the 
spatial distribution of cluster stars, but this cannot be 
stated until the final selection of members has been
completed. According to the previous discussion, it is
clear that the cluster radius is close to the value
$\sim 40$ arcmin and, therefore, this would be the
optimal sampling radius.

\section{Cluster member selection}
\label{sec_members}

From this point we remove any filtering option and get 
all the stars available in the {\it Gaia} EDR3 catalogue. Even 
though the cluster radius seems to be slightly smaller 
than $40$ arcmin (Figure~\ref{fig_eta}) here we use a 
conservative sampling radius of $R_s= 40$ arcmin. With 
this sampling we make a first iteration to determine 
candidate cluster members based on their proper motions, 
parallaxes and photometric properties.

\subsection{Kinematic candidates}

\begin{figure}[t]
\includegraphics[width=\columnwidth]{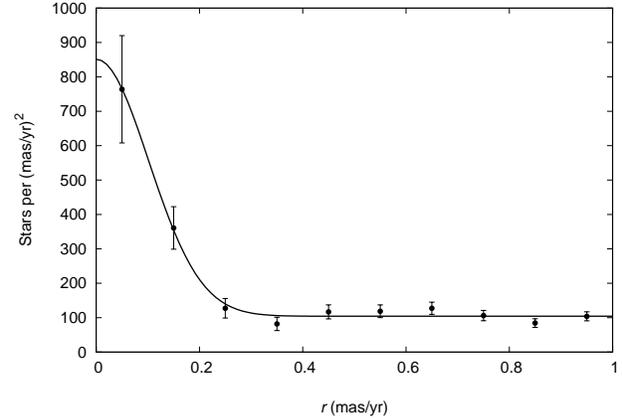}
\caption{Radial density profile of the overdensity in 
the proper motion space centered at $(-1.76,-1.44)$ 
mas~yr$^{-1}$. Used bin size is $0.05$ mas~yr$^{-1}$ 
and error bars are from Poisson statistics. Smooth 
solid line is the best fit to a Gaussian function 
that yields a standard deviation of $0.10$ mas~yr$^{-1}$.}
\label{fig_perfilMP}
\end{figure}

Apart from determining the optimal $R_s$, the used 
algorithm also yields the position of the overdensity 
in the proper motion space. In this first iteration the 
overdensity was found at $(-1.76,-1.44)$ mas~yr$^{-1}$ 
and its radial star density profile
in the proper motion space
is shown in Figure~\ref{fig_perfilMP}.
A Gaussian function with standard deviation 
$\sigma=0.10$ mas~yr$^{-1}$ fitted very well the 
overdensity profile, then we select as kinematic 
candidate members the $125$ stars falling inside 
the $3\sigma = 0.30$ mas~yr$^{-1}$ neighbourhood.

\subsection{Parallactic candidates}

\begin{figure}[t]
\includegraphics[width=\columnwidth]{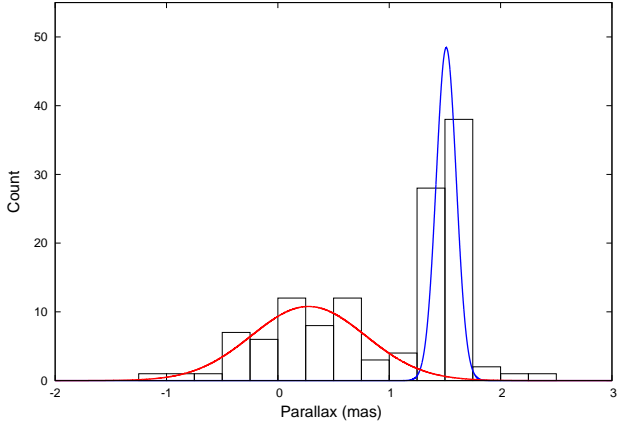}
\caption{Distribution of parallax values for the $125$ 
stars selected by kinematic criteria. Solid lines are 
two Gaussian functions whose sum was fitted to the 
distribution, representing field stars (red line) and 
cluster stars (blue line). The Gaussian corresponding 
to the cluster is centered at $1.512$ mas and has 
standard deviation of $0.090$ mas.}
\label{fig_paralajes}
\end{figure}

Figure~\ref{fig_paralajes} shows the parallax 
distribution for the kinematically selected stars.
We performed a least-square fit to the distribution 
using two (field and cluster) Gaussian functions that 
yielded a cluster mean parallax of $plx=1.512$ mas with 
a standard deviation of $\sigma_{plx} = 0.090$ mas (blue 
line). From here, the parallax selection criterion 
$plx \pm 3\sigma_{plx}$ yields $67$ candidates.

\subsection{Photometric candidates}

\begin{figure}[t]
\includegraphics[width=\columnwidth]{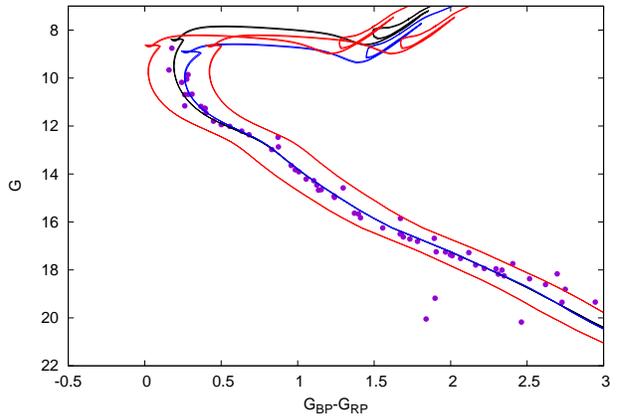}
\caption{Colour-magnitude diagram using {\it Gaia} bands for 
the stars selected by kinematic and parallax criteria 
(blue circles). Solid lines are PARSEC isochrones 
\citep{Bre12} at distance 654~pc, extinction $A_G=0.6$ 
mag and ages of $\log T = 8.4$ Myr (black line) and 
$\log T = 8.6$ Myr (blue line). Red lines represent 
an envelope defined by using the isochrone of age 
$\log T = 8.5$ Myr and shifting it $\pm 0.2$ mag 
along the $(G_{BP}-G_{RP})$ axis, that is used to 
select photometric candidates.}
\label{fig_cmd}
\end{figure}

The colour-magnitude diagram for the $67$ previously 
selected stars is shown in Figure~\ref{fig_cmd}.
The cluster main sequence is clearly visible. In order 
to select photometric candidates we have also plotted 
isochrones generated using PARSEC
v3.4\footnote{http://stev.oapd.inaf.it/cgi-bin/cmd}
\citep{Bre12}.
The distance can be estimated
by adding $+0.017$ mas to the mean
cluster parallax $plx=1.512$ mas for taking into account
the global parallax zero point of {\it Gaia} EDR3
\citep{Gai20}, which gives a distance of
654~pc. Additionally, we assume solar metallicity and 
{\it Gaia} band extinction coefficients given by \citet{Wan19}. 
The best fitted isochrones, determined mainly by eye, were 
those with extinction around $A_G=0.6$ mag and ages in the 
range $\log T = 8.4-8.6$ Myr (black and blue lines in 
Figure~\ref{fig_cmd}). Given that, at this point, we are 
working with only one cluster and few photometric bands, 
it is not necessary to use a more sophisticated fitting 
procedure that will probably lead to similar results 
\citep[e.g.][]{Jef16}.
The obtained age range is smaller than the age given
by \citet{Kha13} ($8.72$ Myr) 
and higher than the recent estimation
by \citet{Dia21} ($8.19$ Myr),
but it is in agreement with the values reported by
\citet{Bos19} ($8.60$ Myr), \citet{Zha19} ($8.45$ Myr)
and \citet{Can20b} ($8.42$ Myr).
As photometric selection 
criterion we use an isochrone of $8.5$ Myr shifted 
$\pm 0.2$ mag in $(G_{BP}-G_{RP})$ (red lines in 
Figure~\ref{fig_cmd}), which yields $60$ stars.

\subsection{Cluster member selection}

\begin{figure}[t]
\includegraphics[width=\columnwidth]{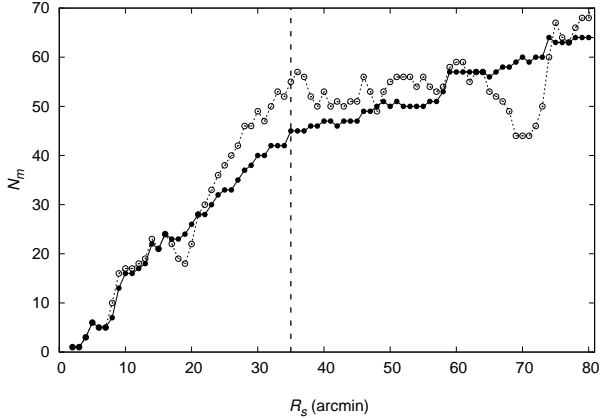}
\caption{Number of estimated members $N_m$ for \at\ as 
a function of the sampling radius $R_s$. Solid circles 
connected by continuous lines refer to members selected 
by using kinematic, parallactic and photometric criteria 
mentioned in Section~\ref{sec_members}, whereas open 
circles with dashed lines are for members based exclusively 
on kinematic criteria. Vertical dashed line indicates the 
radius $R_s=35$ arcmin at which a slope change is observed.}
\label{fig_Rsam}
\end{figure}

Once established the best kinematic, parallactic and 
photometric membership criteria, we repeat the strategy 
of spanning the sampling radius $R_s$. This is the procedure 
originally suggested in \citet{San10} for a reliable estimation 
of the actual cluster radius regardless of the membership 
assignment method. The expected behaviour is that the number 
of members $N_m$ increases as $R_s$ increases until the point 
$R_s = R_c$ and then $N_m$ remains nearly constant (for the 
ideal case of a ``perfect" membership assignment) or increases 
but at slower rate (for more realistic situations with some 
contamination by field stars). Actually, what we do is not 
just to determine the member stars but also to estimate the 
number of spurious members, i.e. the expected number of 
field stars fulfilling the membership criteria. For this,
we use the ``local" field stars defined as those stars 
falling in a concentric ring around the proper motion 
overdensity. The calculated local field density of stars 
allows us to estimate the proportion of field star 
contamination in the overdensity and in the final 
number of assigned members. Then, we estimate $N_m$ 
as the number of stars fulfilling the membership 
criteria minus the expected number of spurious 
members. Figure~\ref{fig_Rsam} is the plot of 
$N_m$ versus $R_s$ for \at\ using data from {\it Gaia} EDR3.
Open circles refer to the number of members estimated by 
using exclusively kinematic criteria (proper motions), 
which is the strategy previously used in 
\citet{San10,San18,San20}. Although the curve is somewhat 
noisy, the expected change of slope is clearly observed 
around $R_s=35$ arcmin. If the full set of membership 
criteria is used (solid circles) then the change of slope 
is less pronounced but still visible at the same point. 
This result is consistent with Figure~\ref{fig_eta} but 
now we can undoubtedly assign the value $R_c=35$ arcmin 
to the cluster radius. According to our results, the 
number of stars fulfilling the membership criteria 
are $55$ from which we expect that $\sim 10$ are 
spurious ($N_m=45$).

\section{Spectral energy distributions}
\label{sec_seds}

The physical properties of the stars can be derived by 
fitting theoretical spectra to their Spectral Energy 
Distributions (SEDs). In order to create a well sampled 
SED (over a wide wavelength range), usually it is assembled 
or complemented with photometric information from different 
catalogues. In these cases, great care must be taken when 
cross-matching different catalogues and combining different 
photometric systems. An advantage of using OAJ data rather 
than cross-matching different catalogues is that we have 
homogeneous multiband photometry in $12$ filters spanning 
the entire optical range. The SEDs of the $55$ selected 
stars were analyzed using the Virtual Observatory SED 
Analyzer (VOSA) tool developed by the Spanish Virtual 
Observatory \citep{Bay08}. VOSA has a friendly interface 
that allows the user to choose different models and 
parameter ranges to find the spectrum that best 
reproduces the observed data.

\begin{figure*}[t]
\includegraphics[width=\columnwidth]{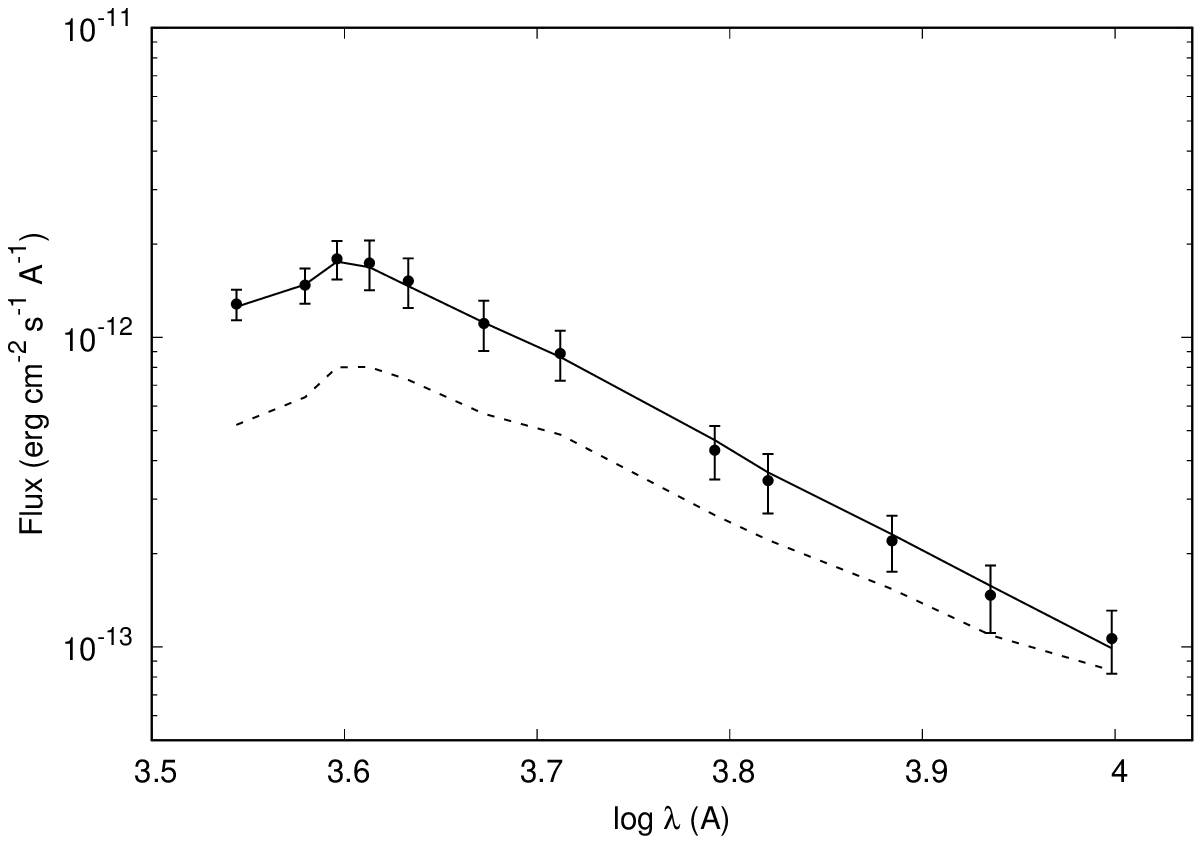}
\includegraphics[width=\columnwidth]{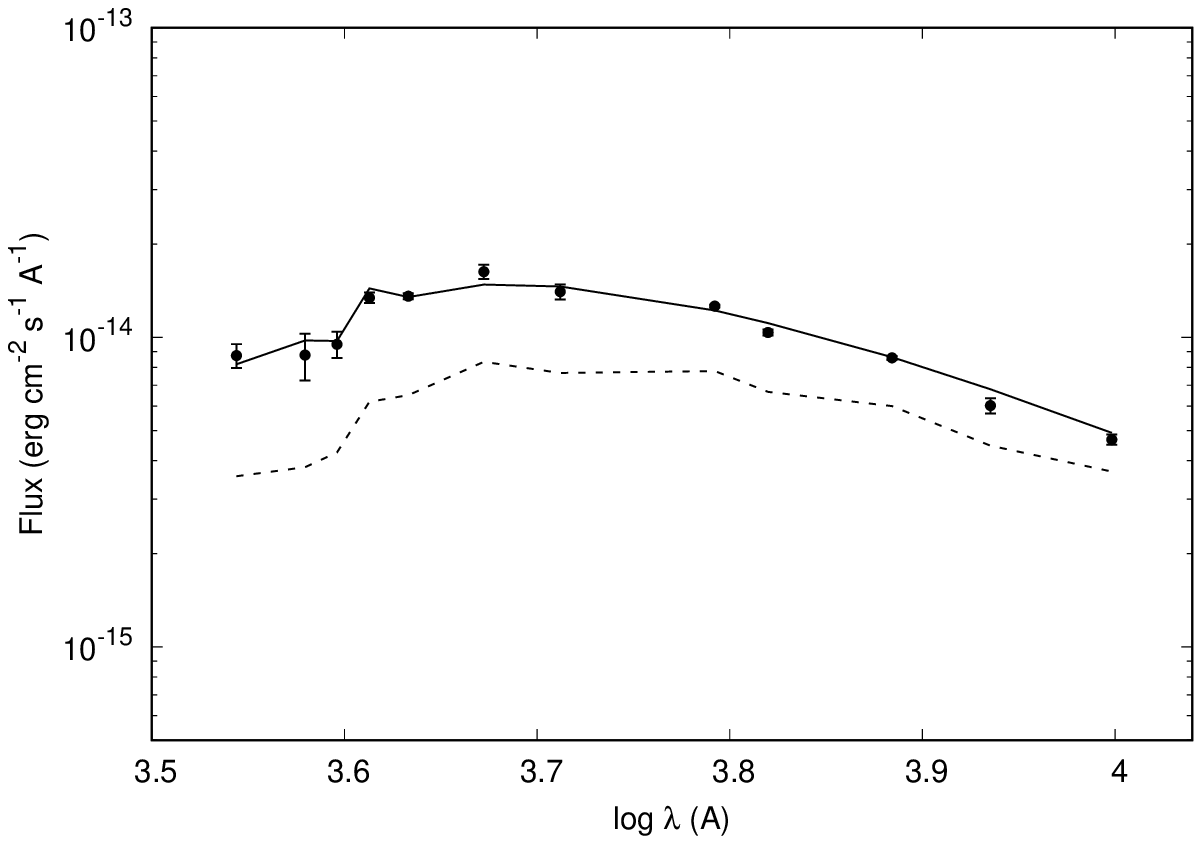}
\caption{Two examples of fitted SEDs. Dashed lines are 
the original observed photometric data, solid circles 
with error bars indicate dereddened data points and 
solid lines are the best fitted Kurucz model. Left 
panel corresponds to the best result for which 
$T_{eff} = 13000$~K and $A_V=0.6$ mag whereas 
right panel is the worst case with 
$T_{eff} = 6000$~K and $A_V=0.6$ mag.}
\label{fig_seds}
\end{figure*}

\begin{figure}[t]
\includegraphics[width=\columnwidth]{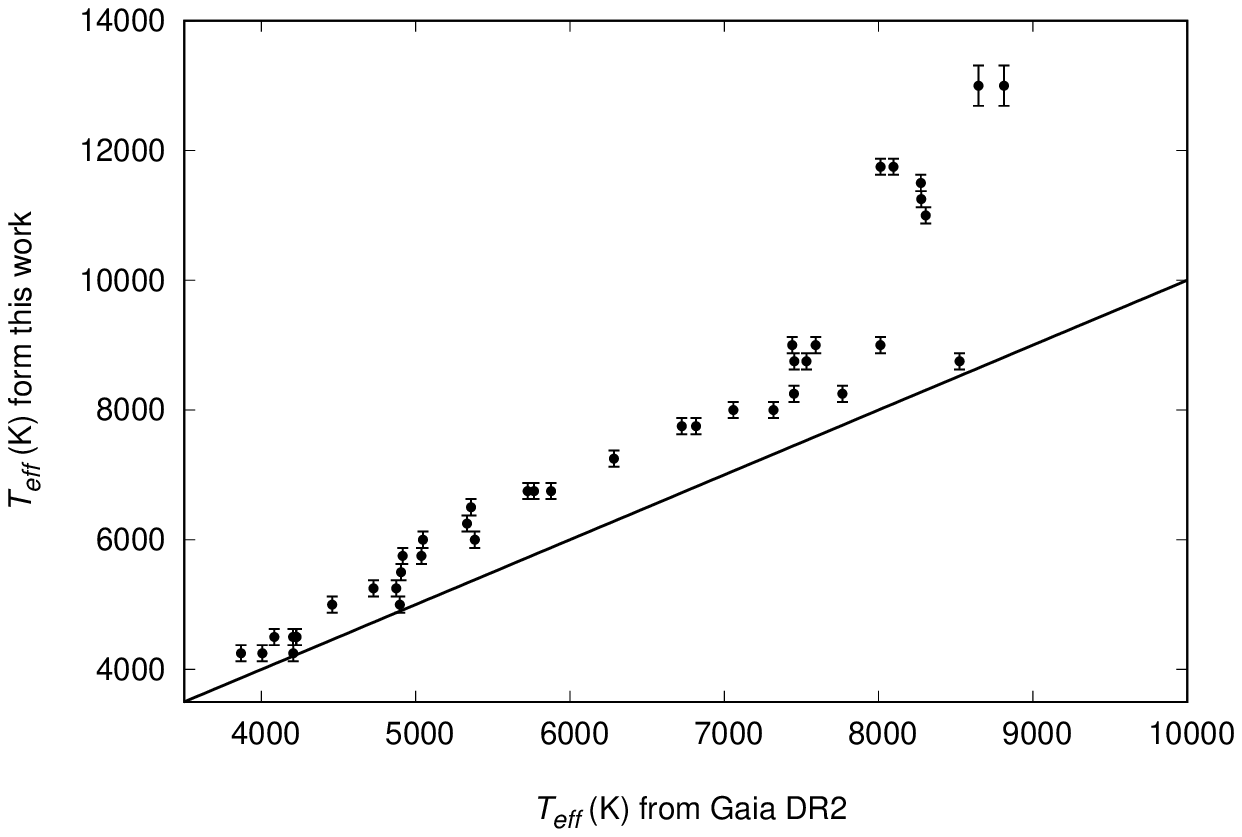}
\caption{Comparison between effective temperatures reported 
in {\it Gaia} DR2 and those derived in this work for the $41$ stars 
with data in common. Solid line is the 1:1 relation.}
\label{fig_Teff}
\end{figure}

For the fitting procedure we used the ATLAS9 Kurucz 
ODFNEW/NOVER models \citep{Cas97}. Given that the fitting 
process is quite insensitive to some parameters, such as 
surface gravity \citep{Bay08}, we fixed both metallicity 
and surface gravity around the expected values, i.e. solar 
metallicity and $\log g = 4$.\footnote{We carried out some 
tests keeping metallicity and surface gravity as free 
parameters and the results did not differ significantly 
from those reported here.} Additionally, we adopted the 
previously obtained distance ($654$ pc) to all the stars. 
Both the effective temperature ($T_{eff}$) and the visual 
extinction ($A_V$) were considered as free parameters. 
Even though not all the stars had valid magnitudes in 
the $12$ photometric bands, their SEDs were well-fitted 
by the models and converged to valid (physically reasonable) 
solutions. Figure~\ref{fig_seds} shows the results for two 
examples: the best and the worst obtained fit.
$T_{eff}$ and $A_V$ values were inferred for
the $55$ stars from the fit of their SEDs.
Effective temperatures are all in the range 
$T_{eff} = 3500-13000$~K. The mean $A_V$ value
is $0.65$ mag (the median is $0.60$ mag) with a
standard deviation of $0.19$ mag. 
Our mean visual extinction is
higher than the value reported by \citet{Bos19}
($0.443$) and smaller than that by \citet{Dia21}
($0.82$), but it agrees with the extinction estimated
by \citet{Can20b} ($0.63$).
There are $41$ stars with 
extinctions and temperatures calculated 
using only {\it Gaia} DR2 photometry by \citet{And18}. 
For these $41$ stars, the mean extinction reported in
{\it Gaia}'s $G$ band is $A_G=0.66$ mag (standard
deviation $0.23$ mag). However, effective 
temperatures reported in {\it Gaia} DR2 are systematically 
smaller than the ones derived in this work by using multiband 
photometry from OAJ, as can be seen in Figure~\ref{fig_Teff}.
The data points that deviate the most from the 1:1 line
in Figure~\ref{fig_Teff}
correspond to the hottest stars ($T_{eff} > 10^4$ K) and 
this is because \citet{And18}'s calculations were restricted 
to the range $3000-10000$ K.

\section{Properties of the open cluster Alessi-Teutsch~9}
\label{sec_properties}

The full list of cluster members, including properties 
resulting from SED fitting, is provided in 
Table~\ref{tab_members} (fully available online).
\begin{table*}[h]
\centering
\caption{List of star members of the open cluster \at\ and
their derived properties including equatorial coordinates 
(J2000), proper motions, parallaxes, magnitudes and colors 
in the {\it Gaia} bands, visual extinctions, effective 
temperatures, bolometric luminosities, and stellar masses 
and radii.}
\label{tab_members}
\begin{tabular}{cccccccccccc}
\hline
RA\_J2000 & DE\_J2000 & $\mu_\alpha\cos\delta$ & $\mu_\delta$ &
$plx$ & $G$ & $BP-RP$ & $A_V$ & $T_{eff}$ & $L_{bol}$ &
$M$ & $R$ \\
(deg) & (deg) & (mas/yr) & (mas/yr) & (mas) &
(mag) & (mag) & (mag) & ($K$) & ($L_{\sun}$) & ($M_{\sun}$) &
($R_{\sun}$) \\
\hline
 51.870058 & 34.983293 & -1.720 & -1.398 & 1.4843 & 13.910 & 1.007 & 0.60 & 6000 & 1.35 & 0.42 & 1.07 \\
 51.915978 & 35.023862 & -1.776 & -1.462 & 1.4968 & 11.943 & 0.501 & 0.60 & 8000 & 8.66 & 0.84 & 1.52 \\
 51.806890 & 34.944837 & -1.857 & -1.639 & 1.5207 & 12.362 & 0.681 & 0.60 & 7250 & 5.72 & 0.83 & 1.51 \\
 51.788046 & 34.984700 & -1.757 & -1.351 & 1.6150 & 19.350 & 2.727 & 0.75 & 3500 & 0.01 & 0.04 & 0.35 \\
 51.909316 & 34.914086 & -1.702 & -1.435 & 1.5637 & 12.470 & 0.871 & 0.60 & 6750 & 5.05 & 0.97 & 1.63 \\
...\\
\hline
\multicolumn{12}{l}{Note: here we show only
a portion for guidance regarding table form and content.
The full version is available online.}
\end{tabular}
\end{table*}
The mean properties of \at\ derived directly from the 
selected members are summarized in Table~\ref{tabAT9}, 
where a comparison is also made with other works, 
especially the most recent studies using data from 
{\it Gaia} DR2 \citep{Bos19,Can20a,Can20b}.
\begin{table}[t]
\caption{Mean parameters of \at}
\label{tabAT9}
\resizebox{\columnwidth}{!}{
\begin{tabular}{@{}lccc}
\hline
Parameter & This work & Other works & Ref.\\
\hline
Right ascension (deg) & $51.8261$ & $51.8700$ & (1) \\
Declination (deg)     & $34.9361$ & $34.9810$ & (1) \\
$\mu_\alpha\cos\delta$ (mas~yr$^{-1}$) & $-1.801$ & $-1.737$ & (1) \\
$\mu_\delta$ (mas~yr$^{-1}$)           & $-1.427$ & $-1.368$ & (1)\\
Number of members & $55$ & $71$ & (1) \\
Distance (pc) & $654$ & $672$ & (1) \\
              &       & $660$ & (2) \\
              &       & $606$ & (3) \\
              &       & $700$ & (4) \\
              &       & $643$ & (5) \\
log(age) (Myr) & $8.5$ & $8.42$ & (2) \\
               &       & $8.60$ & (3) \\
               &       & $8.72$ & (4) \\
               &       & $8.18$ & (5) \\
               &       & $8.45$ & (6) \\
$A_V$ (mag) & $0.65$ & $0.63$  & (2) \\
            &        & $0.443$ & (3) \\
            &        & $0.820$ & (5) \\
Radius (arcmin) & $35$ & $31.8$ & (4) \\
                &      & $29.8$ & (7) \\
                &      & $25.7$ & (8) \\
Total mass ($M_{\sun}$) & $35.8$ \\
Total luminosity ($L_{\sun}$) & $992$ \\
\hline
\multicolumn{4}{p{\columnwidth}}{
References:
(1) \citet{Can20a};
(2) \citet{Can20b};
(3) \citet{Bos19};
(4) \citet{Kha13};
(5) \citet{Dia21};
(6) \citet{Zha19};
(7) \citet{Dia14};
(8) \citet{Sam17}.}
\end{tabular}}
\end{table}
Out of these $55$ members, $45$ are listed in the catalogue 
of \citet{Can20a}, although only $37$ were assigned as 
probable members ($p > 0.5$). 
If we restrict to stars brighter than
$G = 18$ mag, the constraint imposed by \citet{Can20a},
then we get $49$ members from which $45$ ($92$\%) are
in common with their catalogue. The $4$ remaining stars
are fainter than $G = 17$ mag but showing no particularities
in spatial positions, proper motions, parallaxes or their
errors and they likely come from the different methods used
or from the differences in proper motions and parallaxes 
between {\it Gaia} DR2 and EDR3. However,
the sampling 
radius used by \citet{Can20a} is around twice ours. If we 
use $R_s=70$ arcmin instead of $R_s=35$ arcmin we obtain a 
total of $79$ stars fulfilling membership criteria, a number 
very similar to the $71$ members reported by \citet{Can20a}.
In this case, the estimated number of spurious members is $19$, 
which means a field star contamination of $24$\%. Then, it 
is reasonable to assume that about a quarter of \citet{Can20a}'s 
members are actually spurious. Using an unsuitable sampling 
radius may obviously affect the estimate of some important 
parameters, such as for instance the total cluster mass. 
Other parameters could be less affected. For \at, using 
$R_s=70$ arcmin, cluster's center shifted by $\sim 4$ arcmin, 
proper motion centroid by $\sim 0.01-0.02$ mas~yr$^{-1}$ and 
distance remained unchanged. However, depending on the nature 
of the data and on the details of the used method, these 
differences could be more significant and hence the 
importance of a reliable {\it previous} determination 
of the optimal sampling radius.

\begin{figure}[t]
\includegraphics[width=\columnwidth]{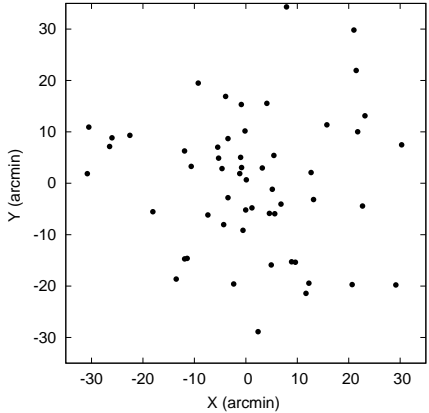}
\caption{Spatial distribution of \at\ members relative 
to the cluster's center.}
\label{fig_XY}
\end{figure}

\begin{figure}[t]
\includegraphics[width=\columnwidth]{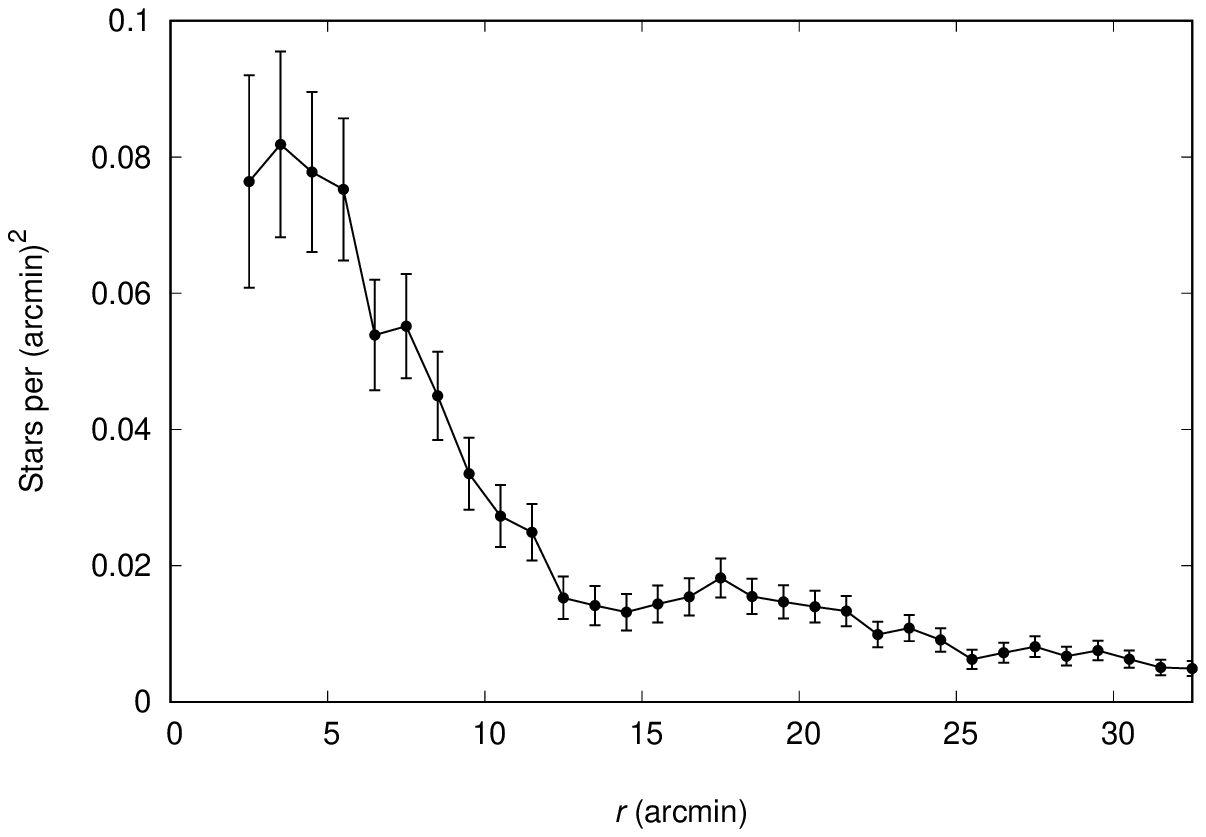}
\caption{Radial spatial density profile for the cluster \at. 
The bins are $5$ arcmin wide and the vertical error bars are 
estimated by assuming Poisson statistics.}
\label{fig_perfilXY}
\end{figure}

Figure~\ref{fig_XY} shows the spatial distribution of 
the $55$ member stars.
It appears quite irregular although a central concentration
of stars can be discerned. The corresponding radial density
profile is plotted in Figure~\ref{fig_perfilXY}.
The small ``valley" seen around $r \sim 15$~arcmin roughly 
agrees with that observed in Figure~\ref{fig_eta} and 
indicates an apparent lack of stars in this region. Given 
the relatively low number of members, this feature is likely 
a random fluctuation. What does seem clear is a two-component 
structure in the radial profile: a more concentrated core of 
radius $R_{core} \sim 12-13$ arcmin and an outer halo extending 
out to $r = R_c$. These values are consistent with the 
relation $R_c \sim 3 R_{core}$ found by \cite{Mac07}.
We analysed separately both populations (core and halo) but 
we did not find patterns or significant differences between them.

Cluster mean proper motion derived in this work
differs by less than $\sim 0.1$ mas~yr$^{-1}$ from that calculated
by \citet{Can20a}, that is the typical proper motion uncertainty
in the sample.
The differences 
with earlier works \citep{Kha13,Dia14,Sam17} are much larger 
($> 1$ mas~yr$^{-1}$) but this is not surprising since these 
studies were carried out with less precise data from the PPMXL 
or UCAC4 catalogues. The number of assigned members in these 
catalogues previous to {\it Gaia} is unrealistically high ($\sim 10^3$) 
likely because of the same reason: larger proper motion errors 
imply larger overdensity areas and a higher number of kinematic 
candidates.

Regarding the cluster distance, our result ($654$ pc) lies in 
between the estimates by \citet{Kha13} and \citet{Bos19} and 
close to the result of \citet{Can20a} (see Table~\ref{tabAT9}). 
At this distance, the cluster linear radius is $6.7$~pc and the 
corresponding number density of stars would be $\Sigma_c \simeq 
0.06$ star/pc$^3$. This makes \at\ one of the less dense known 
OCs. Star density is a parameter difficult to determine accurately 
because it depends on the actual cluster radius, which is an 
uncertain parameter as already discussed in this work \citep[see 
also][]{San18,San20}. We cross-matched \citet{Can20a}'s catalogue 
with de \citet{San20}'s catalogue of reliable OC radii and we 
found $334$ clusters in common, of which only one had 
a star density smaller than \at\ (Ruprecht~45 with $\sim 0.04$ 
star/pc$^3$). The corresponding relaxation time, the time 
necessary for the cluster to reach a Maxwellian velocity 
distribution, can be calculated as\footnote{This equation 
actually refers to the relaxation time {\it at the half-mass 
radius}, because $T_E$ varies through the cluster.} \citep{Spi71}
\begin{equation}
T_E = \frac{0.06 ~ N_m^{1/2} ~ R_h^{3/2}}{G^{1/2}
~ \overline{m}^{1/2} ~ \log (0.4 N_m)}
\end{equation}
being $N_m$ the number of cluster members, $R_h$ the half-mass 
radius and $\overline{m}$ the mean mass of the cluster stars 
($G$ is the universal gravitational constant). Taking 
$\overline{m} = 0.5 M_{\sun}$ and assuming that $R_h 
\simeq R_c/2$ we get $T_E \simeq 40$ Myr, which means 
that \at\ is a (very) relaxed system. Additionally, 
based on the diagnosis criterion proposed by \citet{Gie11}, 
we can say that \at\ is gravitationally bound because its 
crossing time, which is of the same order of magnitude as 
$T_E$ \citep{Spi71}, is significantly smaller than its age.
Note that choosing an unsuitable value of $R_s=70$ arcmin 
(and $N_m=79$) yields $T_E \simeq 130 Myr$. 
Reliable estimates of important characteristic times (e.g. 
dynamical relaxation time, crossing time or evaporation time) 
can be obtained only if reliable measurements for the radius, 
number of members or total mass are available.

\section{Conclusions}
\label{sec_conclusions}

This paper describes a pilot study that implements a novel
approach to determine the properties of an OC in a reliable
way. For a given set of membership criteria, the number of
estimated cluster member should be plotted as a function
of the sampling radius. The point at which the slope of the
curve flattens indicates the actual cluster radius and both
member stars and cluster properties should be derived at
this point. By using this strategy, we used data from
{\it Gaia} EDR3 to study the cluster \at\ and we obtained
a radius of $35$ arcmin and $55$ member stars, from which
we expect that $\sim 10$ are spurious (field stars
fulfilling the membership criteria). We complemented this
data with observed $12$-bands photometry to determine the
final cluster properties (Table~\ref{tabAT9}).
Our results show that \at\ is a bound, dynamically relaxed
cluster having a very low number density ($\simeq 0.06$
star/pc$^3$) and a two-component structure with an outer
halo and a central core of radius $\sim 12-13$ arcmin.

Currently, there is a prevailing approach to determine OC
properties with totally automated techniques 
\citep[see][for recent examples]{Gao18,Yon19,Gao20,Aga21}.
For these cases, we highlight the importance of choosing
an optimal sampling radius {\it previous} to any calculation.
Additionally, it is important to complement the large width
{\it Gaia} bands with photometry in multiple narrow or
medium bands such as the $12$ filters used in this work.
However, addressing this last issue is not
straightforward because it involves dedicated observations
in multiple filters per each cluster.
In this sense, the forthcoming data from the
Javalambre-Physics of the Accelerated Universe Astrophysical
Survey \citep[J-PAS,][]{Ben14} 
could be very useful. J-PAS will cover at least
$8000$ deg$^2$ using a unique set of 56 optical filters
and these data can be used to apply the
proposed approach to OCs falling within the survey area.

\acknowledgments
We are very grateful to the referee for his/her careful
reading of the manuscript and helpful comments and
suggestions, which improved this paper.
NS acknowledges support from the Spanish Ministerio de Ciencia,
Innovaci\'on y Universidades through grant PGC2018-095049-B-C21.
F.~L.-M. acknowledges partial support by the Fondos de 
Inversiones de Teruel (FITE).
Based on observations made with the JAST80 telescope at the 
Observatorio Astrof\'isico de Javalambre (OAJ), in Teruel, 
owned, managed, and operated by the Centro de Estudios de 
F\'isica del Cosmos de Arag\'on. We thank the OAJ Data 
Processing and Archiving Unit (UPAD) for reducing and 
calibrating the OAJ data used in this work.
This work has made use of data from the European Space Agency
(ESA) mission {\it Gaia} (\url{https://www.cosmos.esa.int/gaia}),
processed by the {\it Gaia} Data Processing and Analysis Consortium
(DPAC, \url{https://www.cosmos.esa.int/web/gaia/dpac/consortium}).
Funding for the DPAC has been provided by national institutions,
in particular the institutions participating in the {\it Gaia} 
Multilateral Agreement.

\section*{Funding}
Funding for OAJ and UPAD has been provided by the 
Governments of Spain and Arag\'on through the Fondo 
de Inversiones de Teruel; the Spanish Ministry of Science, 
Innovation and Universities (MCIU/AEI/FEDER, UE) with grant 
PGC2018-097585-B-C21; the Spanish Ministry of Economy and 
Competitiveness (MINECO/FEDER, UE) under AYA2015-66211-C2-1-P, 
AYA2015-66211-C2-2, AYA2012-30789, and ICTS-2009-14; and 
European FEDER funding (FCDD10-4E-867, FCDD13-4E-2685).

\section*{Conflicts of interests}
The authors have no conflicts of interest to declare
that are relevant to the content of this article.

\bibliographystyle{spr-mp-nameyear-cnd}
\bibliography{sanchez_ref}

\end{document}